\documentclass{aa}
\usepackage{graphicx}
\begin{document}

   \title{Letter to the Editor:\\
    On the  Origin of the High Helium Sequence in $\omega$ Centauri}


\author{Andr\'e Maeder, Georges Meynet}

     \institute{Geneva Observatory CH--1290 Sauverny, Switzerland\\
              email:  andre.maeder@obs.unige.ch\\
             email: georges.meynet@obs.unige.ch
               }

   \date{Received  / Accepted }

   \offprints{Andr\'e Maeder} 

   \abstract{The blue Main Sequence (bMS) of $\omega$ Cen implies a ratio of helium to metal enrichment 
   $\Delta Y/\Delta Z \approx 70$, which is a major enigma.
   We show that  rotating models of low metallicity stars, which  account for the anomalous
   abundance ratios of extremely metal poor stars, are also useful for understanding the very high 
   $\Delta Y/\Delta Z$ ratio in $\omega$ Cen. Models of massive stars 
   with moderate initial rotation velocities 
   produce stellar winds with large He-- and N--excesses, but 
    without  the large  C-- (and O--) excesses made by very fast rotation, in agreement with the observed chemical abundance ratios in $\omega$ Cen.  
    
    It is still uncertain whether the abundance peculiarities of  
    $\omega$ Cen result from the fact 
    that  the high velocity contributions of supernovae escaped the globular cluster, usually considered as  a
     tidally  stripped  core of a dwarf galaxy. Another  possibility is a general dominance
     of wind ejecta at very low $Z$, due to the formation of black holes. 
     Some abundance and isotopic ratios like
    $Mg/Al$, $Na/Mg$, $Ne/N$,  $^{12}C/^{13}C$, 
 $^{16}O/^{18}O$ and $^{17}O/^{18}O$ may allow us to further discriminate between these scenarios and between the AGB and massive star contributions.

   }
 \titlerunning{The high Helium Sequence in $\omega$ Cen}   
            
   \maketitle
%

\section{Introduction}

The globular cluster $\omega$ Centauri has remarkable properties: 
it is the most massive globular cluster in the Galaxy and is often interpreted 
as the remaining core of an ancient dwarf  galaxy (Bekki \& Freeman \cite{BekkiF03}), 
a possibility supported by the  density profile of the cluster
(Ideta \& Makino \cite{IdetaM04}). Dynamical studies support a
 formation of $\omega$  Cen
from one of the  small progenitor galaxies of  the Milky Way (e.g. Gnedin et al. \cite{Gnedin02}).

The stars in $\omega$ Cen show a wide spread in metallicity (Norris \& Da Costa \cite{NorrisDC95}) from
$[Fe/H] = -2$ to $-0.5$.
 Among its many  abundance peculiarities, $\omega$ Cen 
 shows a large N--excess, an overabundance of s--elements relatively to Fe
 (Norris \& Da Costa \cite{NorrisDC95}) and  an unusually low $[Cu/Fe]$ ratio 
 relatively to other metal poor stars
  (Cunha et al. \cite{Cunha02};  McWilliam \& Smecker--Hane
 \cite{McWilliam05}), which is interpreted as a relative lack of contributions from supernovae SNIa
  (Cunha et al. \cite{Cunha02}).

The finding of a double sequence in the globular cluster $\omega$ Cen by Anderson (\cite{Anderson97};
see also Bedin et al. \cite{Bedin04}, Gratton \cite{Gratton05}) and the  further interpretation of the bluer sequence by a
strong excess of helium  constitutes a major enigma  for stellar and
galactic evolution (Norris \cite{Norris04}). The interpretation in terms of an He excess is  convincing and supported by stellar models as well by the morphology of the horizontal branch stars (Piotto et al. \cite{Piotto05}).
The great problem is that the bluer sequence with a metallicity
$[Fe/H]= -1.2$  or $Z =  2 \cdot 10^{-3}$ implies an He--content $Y=0.38$ (0.35-0.45), i.e. an He--enrichment
$\Delta Y = 0.14$ (see Norris \cite{Norris04}). 
In turn, this demands a relative helium 
to metal enrichment  $\Delta Y/\Delta Z$ of the order of   70
(Piotto et al. \cite{Piotto05}; Gratton \cite{Gratton05}). 
At the opposite, the
system of globular clusters has a constant  $Y=0.250$ (Salaris et al. \cite{Salaris04}), while $[Fe/H]$
varies a lot, which implies a ratio $\Delta Y/\Delta X(\mathrm{Fe})=0$, where $X(\mathrm{Fe})$ is 
the iron mass fraction.  
The value   $\Delta Y/\Delta Z=70$ is  enormous and more than one order of magnitude larger than the current value of $\Delta Y/\Delta Z= 4-5$ (Pagel et al. \cite{Pagel92}) obtained from  
  extragalactic  HII regions.  A value of 4 -- 5  is  consistent with the chemical yields from supernovae
(Maeder \cite{Maeder92}) forming  black holes above about 20 -- 25 M$_{\odot}$. 

The subject of the present work is to examine whether this extreme value of 
 $\Delta Y/\Delta Z$ can be accounted for by models  of rotating stars at very low 
metallicity. These models, which have the same physical ingredients as the models successfully used at solar $Z$, well account (Meynet et al. \cite{MEMZfirst}) for the abundance anomalies observed in extremely metal poor halo stars.

Sect. 2  collects the relevant observational determinations of the chemical abundances in $\omega$ Cen
and mentions the possible interpretations of  the observed abundance peculiarities.
 In Sect. 3, we show results of  models of rotating stars and in Sect. 4 we compare the results to
  observations of the bMS in $\omega$ Cen. Sect. 5 gives the conclusions. 

\section{The  chemical abundances of the blue Main Sequence (bMS) in $\omega$ Cen }
 
 Let us  quote the anomalies related to the blue Main Sequence   according to Piotto et al. (\cite{Piotto05}):
 
 \begin{itemize}
 \item The bMS is best supported by a strong helium enhancement (Norris \cite{Norris04}). A
  metallicity  $Z= 2 \cdot 10^{-3}$ with a  helium content $Y=0.38$
  is obtained by Piotto et al. (\cite{Piotto05}) by comparing observed spectra with 
  synthetic spectra and adjusting the stellar structure models. For the 
  red Main Sequence (rMS),  these values are $Z=  10^{-3}$ and
 $Y=0.246$. Thus, the  $\Delta Y/\Delta Z$ ratio corresponding to the difference between the rMS and the bMS is
 $\sim 130$.
 \item The metallicity of the bMS is $[Fe/H]= -1.26$ compared to -1.57 for the rMS.
 \item The carbon to metal excess is $[C/M]=0$ for both the bMS and rMS. 
 \item For nitrogen, Piotto et al.
 find $[N/M]=1.0 - 1.5$ for the bMS and $[N/M] \leq 1.0$ for the  rMS.
 \item For  s--elements,  one has for example  $[Ba/M]=0.7$ (with an excess $[Ba/Eu]$) for the bMS and 0.4 for the rMS.
 \item The scatter of the abundances of the bMS stars is small, while it is large for  rMS'.
\item From radial velocity and proper motions studies, there is no difference 
  between stars on the rMS and bMS: both sequences  belong to $\omega$ Cen (Piotto et al. \cite{Piotto05}). 
\end{itemize}

Various possible explanations of the peculiar helium content of the bMS stars
have been mentioned by  Norris (\cite{Norris04}) and by 
Piotto et al. (\cite{Piotto05}). A possibility is galactic winds where the wind of high mass supernovae
escaped the core of the dwarf  galaxy, while the He--rich winds of lower mass SNII are kept in.
The models with mixing and fallback (Umeda \& Nomoto \cite{Umeda03}) offer a wide range of possibilities 
due to the internal freedom in the choice of parameters, however the agreement with observational constraints has not yet been demonstrated. Piotto et al. (\cite{Piotto05}) examine the possibility that the required helium  is produced by AGB stars, however they note that the amount of helium produced seems  insufficient. 
In addition, AGB  models would also produce C efficiently and this is not observed.
Norris (\cite{Norris04}) proposed massive stars as the source  of helium.
However, the yields of massive stars, as mentioned above, usually lead to
 much lower  $\Delta Y/\Delta Z$ ratios than the one necessary for
   $\omega$ Cen. Another  possibility mentioned  by Piotto et al. (\cite{Piotto05}) is
 that $\omega$ Cen was enriched by supernovae in the mass range of 10--14  M$_{\odot}$, which
 produce significant He as well as some N
(Thielemann et al. \cite{Thielemann96}). The problem is, however, that such stars are not the favored source of s--element and it is hard to understand why the contributions of more massive stars are absent.
Bekki and Norris (\cite{BekkiNorris05}) suggests that the bMs is formed from gas ejected from field stars
that surrounded $\omega$ Cen, when it was a galaxy nucleus. However, the star models  used to draw this conclusion do not account for rotationally enhanced mass loss.

\section{Rotating models of very low metallicity  stars}

The main   features of rotating  low $Z$ star models are 
the following ones  (Maeder \& Meynet \cite{MMVII};  Meynet et al. \cite{MEMZfirst}):

\begin{itemize}
 \item Low $Z$ stars easily reach break--up during  MS phase and thus lose mass, by the combined effect
 of  stellar winds and rotation (cf. Maeder \& Meynet \cite{MMVI}).
\item Low $Z$  stars are   compact and the external Gratton--\"{Opik} cell 
of  meridional circulation is negligible due to the high stellar density. This leads to 
   steep  $\Omega$--gradients in both AGB and massive stars at very low $Z$, which
 favor  mixing of the chemical elements from the He, C, O core
 into the H--burning shell. 
\item The mixing produces large surface enrichments in nitrogen and, depending on rotation, in other heavy elements like C, O, Ne and s-elements.
 This allows  radiative stellar winds, so that  low $Z$ stars  experience   mass loss
during their AGB or red supergiant phases. 
\item If the star makes a blue excursion in the HR diagram after the red supergiant stage, the contraction
of the convective envelope brings it to critical rotation and drives intense winds.
\end{itemize}

\begin{figure}
\resizebox{\hsize}{!}{\includegraphics[angle=00]{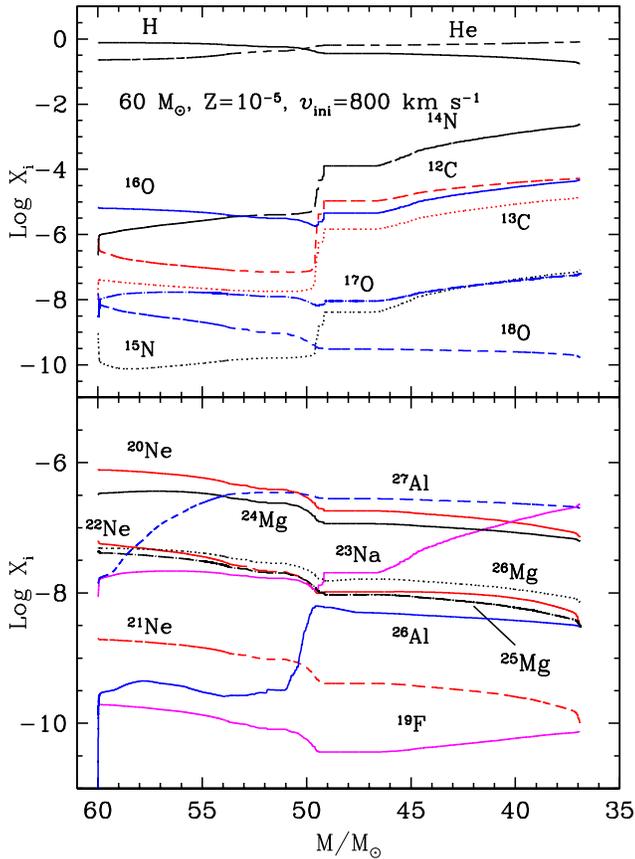}}
\caption{Evolution as a function of the remaining mass of the surface 
abundances in mass fraction for a 60 M$_\odot$ 
stellar model at $Z=10^{-5}$ and with $\upsilon_{\rm ini}=800$ km s$^{-1}$, i.e.
 an initial ratio  of angular velocity to the critical value $\Omega/\Omega_{\mathrm{c}}=0.85$ .}
\label{xisurf}
\end{figure}

\begin{figure}
\resizebox{\hsize}{!}{\includegraphics[angle=-90]{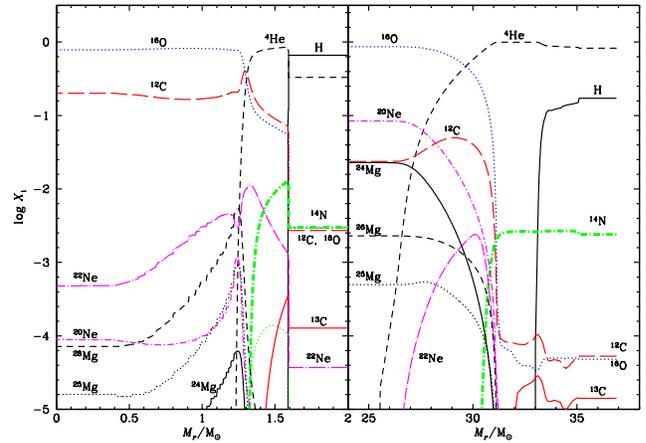}}
\caption{Variations of the abundances (in mass fraction) as a function of the Lagrangian mass
for  (left) a 7~M$_\odot$ stellar model in the early AGB phase, and for (right) a 60~M$_\odot$ 
 model at the end of the core
C-burning phase. Both models were computed
with $\upsilon_{\rm ini}$=~800~km~s$^{-1}$ corresponding respectively to $\Omega/\Omega_{\mathrm{c}}=0.83$
and $0.85$. Both models have $Z=10^{-5}$.}
\label{evolxi}
\end{figure}

We examine  the problem of $\omega$ Cen  in relation with
   models of massive and  AGB--stars. Some models were already calculated  (Meynet et al.
\cite{MEMZfirst}) and we also give here a  model of massive stars at $Z= 10^{-5}$ 
for  an intermediate rotation velocity.
The results are shown in  Figs. \ref{xisurf} and \ref{evolxi} and    in Table \ref{tblwind}.

Fig. \ref{xisurf} shows the evolution of the surface abundances as a function of the remaining mass 
(which is equivalent to an age scale) for a model with an initial mass of 60 M$_\odot$ and Z=$10^{-5}$.
The initial fraction of the break--up angular velocity is $\Omega/\Omega_{\mathrm{c}}=0.85$, which corresponds  to $\upsilon_{\rm ini}$= 800 km s$^{-1}$. (It is likely that the fraction of the break--up velocities
 is the meaningful quantity  to consider rather than the $v \sin i$, since during star formation
 there is  an enormous excess of angular momentum to be dissipated to allow the formation of star with sub critical velocities.)
 During a first phase, the actual mass 
decreases from 60 to about 51 M$_\odot$. The changes of the surface abundances are due to rotational mixing 
 of CNO processed material (mild decreases of $^{12}$C and
$^{16}$O  and increase of $^{14}$N, the sum of CNO elements remaining constant during this phase).
The mass lost, of about 9 M$_\odot$, results from radiatively driven stellar winds and 
evolution at the break-up limit.

When the actual mass is $\sim$51 M$_\odot$, the star is at the middle of the core He--burning phase 
($Y_{\rm c}$ equal to 0.45) and with log $T_{\rm eff}$=3.850. An outer 
convective zone  deepens in mass, dredging-up material
to the surface. This produces  sharp increases of the surface abundances in $^{14}$N, $^{12}$C,
$^{13}$C, $^{15}$N as well as in Na and Al. The enrichment in oxygen is  modest.  The 
amount of heavy elements increases up to more than 240 times the initial $Z$.  From
 $\sim 50$ M$_{\odot}$, the star is essentially an He--star,  corresponding to a Wolf--Rayet of type
 WN. A total of  12.20 M$_{\odot}$ of helium is ejected, of which 5.86 M$_{\odot}$
  of newly synthesized helium (see Table \ref{tblwind}).

Fig. \ref{evolxi} compares the structures of an initial 7 M$_{\odot}$
in the early AGB phase and of the previous  
60 M$_{\odot}$  model at the end of the central C--burning phase, the initial $\Omega/\Omega_{\mathrm{c}}=0.83$ and 0.85 respectively.  
The 7 M$_\odot$ model, in contrast with the 60 M$_\odot$ model, does not reach the break-up limit during the MS phase, however it shows more internal mixing. This is still 
visible in  advanced stages (Fig. \ref{evolxi}),  where one notices  milder composition
gradients at the edge of the core in the AGB model of  7 M$_{\odot}$ than in the late phases of the 60 M$_{\odot}$ model.
This  results from the steeper internal $\Omega$--gradient in the 7 than in the 60 M$_{\odot}$.
The origin of this difference is the so--called Gratton-\"{O}pik cell of meridional circulation, which brings 
angular momentum outward. This circulation term is weaker when density is higher (e.g. Maeder \& Meynet \cite{MMVII}), this is what happens in the 7 compared to the 60 M$_{\odot}$ models at Z=10$^{-5}$: less angular momentum is 
evacuated from the stellar center, henceforth  steeper $\Omega$--gradients and more mixing.

The 7 M$_\odot$ model at $Z=10^{-5}$ remains in the blue part of the HR diagram during the whole He-burning phase,
preventing an outer convective zone to appear and to dredge-up the primary CNO elements.
Only at the end of the core He-burning phase, it evolves to the red and approaches
the base of the Asymptotic Giant Branch. An outer convective zone appears
and produces an enormous enhancement of the surface CNO elements (cf. Fig.~\ref{evolxi}). In contrast, the weaker mixing of the 60 M$_{\odot}$ model
does not bring oxygen and carbon to a level comparable to that of $^{14}$N, which as shown by 
Fig. \ref{xisurf} keeps all the way higher. We also see the high helium  content at the surface of
the 60 M$_{\odot}$ model as a result of mass loss and mixing, while the He--enhancement is  modest
in the 7 M$_{\odot}$ AGB model.

\begin{table}
\caption{Comparison of the   wind ejecta  of a  60 M$_{\odot}$  and
a 7 M$_{\odot}$  model with $Z=10^{-5}$ for  different values of $\Omega_{\mathrm{c}}$. 
$\Delta Y$ and $\Delta Z$ are the amounts of new 
helium and heavy elements ejected. The brackets indicate the ratios in log scale 
of the abundances in the winds or in the AGB envelopes compared to the solar ratios.} 
\label{tblwind}
\begin{center}
\begin{tabular}{lccccccc}
 $M$                  & 60 M$_{\odot}$    & 60 M$_{\odot}$   & 7 M$_{\odot}$   \\  
 $\Omega/\Omega_{\mathrm{c}}$    & 0.85  &  0.38  & 0.83 \\
                      &                   &                  &                 \\
\hline
                         &                               &                &                 \\
$\Delta Y$ in M$_{\odot}$& 5.86                          & 1.73           &   0.69          \\ 
$\Delta Z$ in M$_{\odot}$& 0.09                          & 2.6e-05        &   0.12          \\
$\Delta Y/\Delta Z$      &  63.3                         & ~1e+05         &   5.69          \\
$[C/Fe]$                 &   3.05                        & -0.55          &   4.25           \\
$[N/Fe]$                 &   4.36                        &  1.79          &   4.70           \\
$[O/Fe]$                 &   2.92                        &  0.33          &   3.77           \\
$[N/C]$                  &   1.31                        &  2.34          &   0.45           \\
$[N/O]$                  &   1.44                        &  1.43          &   0.93           \\
$[Ne/N]$                 &  -3.90                        &                &  -1.90           \\
$[Ne/Na]$                &  -0.80                        &                &   0.78          \\
$[Na/Mg]$                &   1.00                        &                &   1.26          \\
$[Mg/Al]$                &  -0.90                        &                &   0.193        \\
$^{12}C/^{13}C$          &   3.81                        &  7.06          &   76.97        \\
$^{16}O/^{18}O$          &  1183                         &  2600          &   4200         \\
$^{17}O/^{18}O$          &  2e+06                        & 19000          &   181          \\
                         &                               &                &                \\
\hline
\end{tabular}
\end{center}
\end{table}

The winds  of the models considered  are shown in  Table \ref{tblwind}, one notices the following features:
 \begin{enumerate} 
\item The winds of low $Z$ massive rotating stars eject large amounts $\Delta Y$ of new helium, while   
 the envelopes of AGB, which is at the end entirely ejected, contain only a moderate amount of helium. $\Delta Y/\Delta Z$ of the order of $10^2$, or even more, are produced in massive stars. 
\item There are large excesses of N ejected  for both massive and AGB stars. The N--excess is larger for
higher rotation.
\item Rotating  AGB stars also have high productions of C and O, about  an order of magnitude 
below the N--production. 
\item The winds of massive stars show  some C and O contributions, however even for fast rotation they are smaller than for N by one to two orders
of magnitude. At moderate rotation, the production of C and O are negligible.
\item The winds of massive stars produce  large amounts of  $^{13}$C and moderate  amounts of Na and Al.
\item AGB winds produce    large
excesses  of F, Ne and Na  (cf. also Meynet et al. \cite{MEMZfirst}) and moderate ones in Mg and Al.
\item The large excess of $^{22}$Ne together with  excesses of $^{25}$Mg and
$^{26}$Mg in AGB stars  imply a  significant production of s--elements. 
\item The production of s--elements 
in massive stars is possible if they  lose enough mass to enter the WC stage.
\end{enumerate}

There are evidently important  uncertainties, such as  the rotational velocities of the first stellar generations,  the mass loss rates by stellar winds at low $Z$, the relative importance of the magnetic field,  the distribution of  stellar masses, etc.

Contrarily  to simple expectations, AGB models exhibit
nucleosynthetic products with signatures of higher temperatures than massive stars. In both cases,
 the synthesis is made by the CNO,
NeNa and MgAl  cycles in the H--shell burning, which receives  some
$^{12}$C diffusing by rotational mixing out of the He--burning core. 
In AGB stars, most of the ejection 
occurs rather late in the nuclear evolution, i.e. near the end of the He--burning phase in the TP--AGB
phase. In massive stars, most of the wind ejecta occurs earlier, from a synthesis at lower temperatures $T$
in the course of the He--burning
phase. This  explains the different $T$--signatures, which offer a powerful way to discriminate whether AGB or massive stars are the main source
of the peculiar bMS abundances.

\begin{figure}
\resizebox{\hsize}{!}{\includegraphics[angle=0,width=10cm,height=6.5cm]{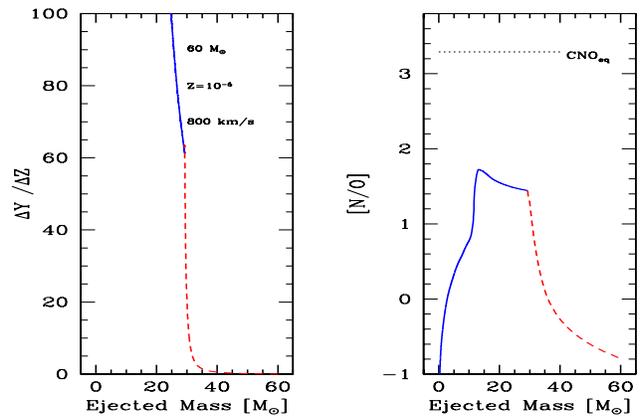}}
\caption{Variations for a 60 M$_\odot$ model with initial $\Omega/\Omega_{\mathrm{c}}=0.85$  and
Z=$10^{-5}$: left) the ratio  $\Delta Y/\Delta Z$  as a function of the ejected mass; right) the same  for the ratio of nitrogen to oxygen normalized to solar.
The continuous line corresponds to the mass ejected by stellar winds, the dashed line shows how the ratios would vary if some amount of  mass from the pre--SN model is added to the ejecta.
The dotted lines indicate the CNO equilibrium values  obtained in the convective core of the star, 
when the mass fraction of hydrogen at the center is 0.34. }
\label{dydz}
\end{figure}

We  examine  how the ratios $\Delta Y/\Delta Z$ and, for example, N/O (in mass fraction) vary as a function
of time or  of the mass ejected in the wind during the stellar lifetime.
This is shown in Fig.~\ref{dydz} for the case of the fast rotating 60 M$_\odot$ stellar model at $Z=10^{-5}$.
At the beginning $\Delta Y/\Delta Z$ is not defined, 
then it tends to infinity when 
some newly synthesized helium appears at the surface
without  heavy elements. 
When the ejected mass approaches about 25 M$_\odot$, $\Delta Y/\Delta Z \sim 100$, 
and at the end of the stellar lifetime $\Delta Y/\Delta Z \sim 63$ in the wind ejecta. 
The new heavy elements  in the winds are in the form of new CNO elements, without new iron.

The dashed line in Fig.~\ref{dydz} shows how the addition 
of material from the pre--supernova (SN) model changes the ratios. 
As soon as some SN ejecta are present, much lower 
values of $\Delta Y/\Delta Z$, N/C and N/O are obtained. 
Even the addition of  small amounts of 
mass ejected by SN changes a lot the ratios of the ejecta. This is a critical point, 
since one does not know  the SN properties and  nature of remnants at low $Z$. If black
holes form, they likely swallow most of the $\alpha$--rich layers of the pre--SN.
In this respect, the $\Delta Y/\Delta Z$ ratio offers a particularly sensitive test of
the relative importance of wind and SN ejecta, the same is true for the ratio of N to
$\alpha$--rich nuclei.


\section{Interpretation of the bMS properties}


We see from Table \ref{tblwind} that
  models of massive stars are able to produce high or even very high $\Delta Y/\Delta Z$ ratios as indicated by the bMS sequence of $\omega$ Cen, a moderate  rotation producing higher ratios. The absence of a
  C--excess points in favor of a not too extreme rotation, maybe of the order of 
 $\Omega/\Omega_{\mathrm{c}}=0.5$.  The observed N-excess, which are not as extreme as in the wind of 
 the fast rotating model, is also in better agreement with an average rotation of the indicated order.
 This is remarkably converging. 
  
 The comparisons  point toward
 a source of the large helium production in massive stars, rather than in AGB stars. This is in agreement
 with the general view that stars of higher masses form at very low $Z$.
 Another  clear conclusion is (cf. Fig. \ref{dydz})  that there is very little contribution to the yields from 
 supernovae, otherwise $\Delta Y/\Delta Z$ would be strongly reduced.
 The above results do not exclude some contributions from AGB stars,  which would not reduce too much the  $\Delta Y/\Delta Z$ ratios and would make some
 contributions to s--elements, however as shown by Piotto et al. (\cite{Piotto05})
 it is  unlikely that AGB stars  can produce enough helium. Another
 possible source of  s--elements is in  massive stars, 
 if they lose enough mass to enter the WC stage. At this stage, the relative contributions of massive and
 AGB stars are unknown.
 We emphasize that further pertinent
 tests on this critical question can be provided by  abundance ratios
 like $Mg/Al$, $Na/Mg$, $Ne/N$ as well as by isotopic ratios such as $^{12}C/^{13}C$, 
 $^{16}O/^{18}O$ and $^{17}O/^{18}O$  which are $T$--sensitive (cf. Table \ref{tblwind}).
 
 The enrichment of $\omega$ Cen seems to result mainly from stellar winds, 
 without the usual contributions in heavy elements from supernovae.  What is the reason for that ?
 One may envisage two possibilities, which are not mutually exclusive:
 
 \medskip
 \noindent
 A)  The particular chemical history of $\omega$ Cen:
 
 \smallskip
 \noindent
 $\omega$ Cen likely  experienced a high intial rate of star formation and a fast initial chemical enrichment. This hypothesis is  consistent 
 with both the absence of extremely  metal poor stars with $[Fe/H] \leq -2$ and the evidence by Cunha et al. (\cite{Cunha02})  of the  absence of contributions from SNIa up to $[Fe/H]= -0.8$. 
 
 Now, the lowest  mass
  stars of the bMS down to at least 0.2 M$_{\odot}$, as observed in $\omega$ Cen, needs about $2 \cdot 10^8$ yr to be formed since the time they left the birthline (cf. Stahler \& Palla \cite{StahlerP04}). This 
  implies that all stars with an initial mass above, say, about 4 M$_{\odot}$ had the time to contribute 
  to the  cluster enrichment. The wind ejecta  of  supergiants and AGB, with
  velocities of less than a few 10$^2$ km s$^{-1}$,  may not escape the cluster core.
  Since  $\omega$ Cen is probably the stripped core of an
   ancient dwarf galaxy feeding the Milky Way (cf. Bekki \& Freeman \cite{BekkiF03}; Gnedin et al. \cite{Gnedin02}), the tidal effects were large enough to remove the external galactic layers. 
   Thus, it would not be surprising that most supernova ejecta at velocity above $10^4$ km s$^{-1}$
    also escaped from the globular core  (cf. also 
   Norris \cite{Norris04}; Piotto et al. \cite{Piotto05}). The mentioned absence of ejecta from 
   SNIa, which may originate from stars initially less massive than 8 M$_{\odot}$ may be an
   additional argument for the escape of the winds from supernovae.
   
   \medskip
   \noindent
   B) The general dominance of enrichments by stellar winds at low $Z$:
   
   \smallskip
   \noindent
   Another attractive hypothesis is that for some range of low $Z$ values, 
   the enrichment by stellar wind is 
   generally  dominant, with reduced  contributions from the onion skin layers of heavy 
   elements in supernovae of type II.
   This would be the case if the core collapse in supernova explosions lead to the formation of black holes, which swallow most of the heavy elements formed. This may occur at lower $Z$, since mass loss, although present,  is not extreme, which thus leads to larger masses in  pre-SN models, which favors black hole formation.

   It is premature to choose between these two possibilities, which are not  exclusive 
    in the case of $\omega$ Cen. Further
   observations of abundance ratios in $\omega$ Cen and their comparison with more extended grids of models
   are the way to follow.

\section{Conclusions}

The wind contributions of low $Z$ massive rotating stars  are able to produce the high $\Delta Y/\Delta Z$
 observed in the  bMS sequence in  $\omega$ Cen.  The observations tend to favour an origin of the high
 helium observed by 
contributions from massive stars of intermediate rotation velocities. 

At this stage, it is not clear whether
the dominance of wind contribution is a general feature of low $Z$ stars or whether 
the tidal
evaporation  experienced by $\omega$ Cen has   enabled it to lose
most of the supernovae ejecta, keeping the enrichment from stellar winds.

Some critical abundance and isotopic ratios may offer further signatures of the 
contributions of AGB and massive stars.



\end{document}